# *Ab-initio* Investigation on *h*-$Be_3N_2$ Monolayer for Photocatalytic Hydrogen Evolution Reaction and Oxygen Evolution Reaction

Karan Patel, Hetvi Jadav, Himanshu Pandey*

*Condensed Matter & Low-Dimensional Systems Laboratory, Department of Physics, Sardar Vallabhbhai National Institute of Technology, Surat-395007, Gujarat, India*

*Corresponding Author E-mail: hp@phy.svnit.ac.in

# 1 Introduction

There has been an unprecedented rise in the demand for energy across the world dues to a quick increase in the world population together with continuous urban development, and industrial growth [1,2]. At present, the majority of energy supply is provided by the conventional sources like fossil fuels, petroleum, and coal [3,4]. Therefore, some typical environmental problems such as greenhouse gas emission, worsening air quality, and fluctuation in the Earth's temperature were created by their continuous consumption [5]. Additionally, the current exhaustion of traditional energy sources due to its limited availability highlights pressing necessity of safe, environmental being, economically viable and sustainable energy solution [6]. Among the multiple energy carriers, green hydrogen stands out as one of the most intriguing possibilities because of its high gravimetric density, abundance, its ability to produced zero carbon emission during its combustions [7,8]. The idea of using hydrogen as clean and renewable energy sources, first proposed by Juels Verne, a French author of science fiction, in his 1874 novel 'The Mysterios Island', in which he stated that 'water will be the coal of the future', indicating that water would ultimately serve as an endless energy resource [9]. Almost one hundred years later, a significant scientific advancement was accomplished by Akira Fujishima and Kenichi Honda in 1972. They effectively produced hydrogen and oxygen through photoelectrochemical splitting of water employing a $TiO_2$ photoanode when exposed to ultraviolet light [10,11].

Inspired by the success of Fujishima and Honda, several techniques for water splitting have been developed to generate hydrogen and oxygen, including electrocatalytic, photocatalytic, photo electrocatalytic, photovoltaic-electrocatalytic, and thermochemical water splitting [12–14]. Among them, semiconductor-based photocatalytic water splitting has drowned a lot of interest from researchers and energy sectors, since it directly converts abundant solar energy into chemical energy (hydrogen) without zero carbon emission [15–17]. In this process, the hydrogen evolution reaction (HER) and oxygen evolution reaction (OER) occurs simultaneously, resulting in the sustainable production of hydrogen from water under sunlight. However, achieving effective complete photocatalytic water splitting is still a challenging task, because of slow OER speed, not absorbing sufficient visible light, and needing costly Nobel metal catalysis [18–20]. Therefore, a lot of studies are still ongoing to create good, long-lasting, and cheap materials that help speed up the reaction needed for splitting water into hydrogen and oxygen, making the whole process more effective.

In the beginning investigation of photocatalysts, three-dimensional (3D) semiconductor materials such as $TiO_2$, ZnO, $WO_3$, CdS, $BiVO_4$, and $Fe_2O_3$ have been extensively researched as photocatalysts for generation of hydrogen from water. Although, these materials exhibit good photocatalytic efficiency, their practical utility is often limited by factors such as poor visible light adsorption due to their wide band gap, limited surface area, and low availability of active catalytic sites [21–26]. The advent of two-dimensional (2D) material has opened new avenues to overcome these limitations. Recently, various forms of graphene, graphitic carbon nitride (g-$C_3N_4$), $MoS_2$, $WS_2$, phosphorene, hexagonal boron nitride (*h*-BN), MXenes, transition metal dichalcogenides as well as various group III–V monolayers have shown excellent catalytic efficiency for HER, OER and complete water splitting [27–30]. Additionally, Be-based two-dimensional materials have gained increased interest due to several benefits such as accessibility, durability, toughness, eco-friendliness, and compatibility, with a significant number of electronic devices relying on these substances [31]. Therefore, 2D materials are considered as highly promising candidates for future solar-driven hydrogen production technologies.

Very recently, several studies on novel Be-based hexagonal 2D materials namely h-$BeN_2$, $Be_2C$, $BeCN_2$, BeO, $Be_2PI$ etc. have been designed through first-principles analysis. Theses discovered beryllium-derived two-dimensional (2D) materials have been shown amazing structural, electrical, and optical properties that make them great candidates for photocatalytic water dissociation. [32–36]. Beryllium's light weight results in materials with a low density and a sizable specific surface area, therefore exposing many active sites that catalyse OER and HER. Furthermore, the powerful covalent contacts between beryllium and surrounding atoms offer remarkable thermal and mechanical resistance, therefore assisting to maintain structural integrity over prolonged photocatalytic activity. Several beryllium-based 2D materials have suitable band gaps in the visible or near-ultraviolet spectrum, which allows for efficient solar energy absorption as well as advantageous placement of valence and conduction band edges that satisfy the thermodynamic requirements for full water splitting. Furthermore, their structural and electronic configuration can be carefully adjusted by employing various strategies, including strain engineering, surface modification, and defect management, to improve charge carrier mobility, tight utilization, and reduction and oxidation efficiency [37]. Moreover, their extremely thin structure substantially reduces the distance that carrier need to cover from the point of generation to surface active's sites, which aids in diminishing electron-hole recombination and improving photocatalytic effectiveness. In addition, the diverse chemical composition of materials based on beryllium allows for the inclusion of multiple non-metal elements, which results in the creation of various stable two-dimensional semiconductors with customized photocatalytic characteristics. These advantages together make beryllium-based 2D materials especially appealing for environmentally friendly solar-powered hydrogen production and inspire continuous theoretical and experimental study on their photocatalytic efficiency.

In this study, we methodically explore the capabilities of the 2D hexagonal $h$-$Be_3N_2$ mono-layer for complete water splitting via HER and OER utilizing density functional theory (DFT) computations. Initially, the structural characteristics are analysed by performing geometry optimization to ascertain the refined lattice parameters, bond lengths, and bond angles, followed by an evaluation of the electronic characteristics using the band structure and projected density of states. The stability in terms of dynamics and temperature of the $h$-$Be_3N_2$ monolayer is additionally assessed using phonon dispersion calculations and ab initio molecular dynamics (AIMD) simulations. To evaluate its capability as a photocatalyst, the positions of the band edges are identified and compared with the redox potentials for water splitting, determining the thermodynamic viability of the HER and OER with solar illumination. Additionally, the catalytic efficiency is analysed by computing the adsorption energies of the intermediates in the reactions and the respective Gibbs free energy profiles for both HER and OER in dark (light off) and illuminated (light on) settings. The impact of pH on the thermodynamics of the reactions is also studied by assessing the pH-sensitive Gibbs free energy, ensuring a thorough understanding of the photocatalytic properties of the *h*-$Be_3N_2$ monolayer for effective overall water splitting.

# 2 Computational Methodology

All computational studies in this research were done using density functional theory with Quantum Espresso code [38,39]. The interaction between core and valence electrons was described with the projected augmented-wave (PAW) method for structural relaxation and electronic parameter calculations [40]. The exchange-correlation interaction among electrons was modelled through the generalized gradient approximation (GGA) with the Perdew-Burke-Ernzerhof (PBE) functional [41]. To presents electronic wavefunctions, a plane wave basis set was used, with a kinetic energy cutoff of 50 Ry and a charge density cutoff of 450 Ry. All atomic configurations were fully relaxed until the total energy and forces on each atom were below $10^{-6}$ eV and $10^{-4}$ eV/Å, respectively. A vacuum region of 22 Å added along the z-axis to limit interaction between successive layers. The dispersion effects related to the 2D nature of the system were captured by including the van der Waals (vdW) interactions using the DFT-D2 correction method developed by Grimme in all calculations [42]. The Brillouin zone was sampled through the Monkhorst-Pack scheme with k-point meshes of 6×6×1 for structural relaxation and 10×10×1 for electronic properties calculations [43]. Ab-initio Molecular Dynamics (AIMD) simulations were performed at a temperature of 300 K for 6 ps using the NVT ensemble with an ultrasoft pseudopotential [44,45].

# 3 Results and Discussion

## 3.1 Structural stability

Figure 1 illustrates the energy-minimized structure of the 3×3×1 *h*-$Be_3N_2$ monolayer, which exhibits a hexagonal lattice with a planer honeycomb-like atomic arrangement. The space group of *h*-$Be_3N_2$ is p6/mmm (No.199). A dashed black line at the center of the figure 1(a) delineates the primitive unit cell, which consist of five (three Be and two N) atoms, whereas the solid dark line outlines the *h*-$Be_3N_2$ supercell. The atoms of beryllium (Be) and nitrogen

(N) are depicted in green and blue color. The optimized lattice parameter is found to be a=b=5.24 Å. In this configuration, each N atom is coordinated with the three nearest-neighbor Be atoms, with a calculated Be-N bond length ($d_{\mathrm{Be-N}}$) of 1.51 Å. The distance between Be-Be and N-N atoms are about 1.62 Å and 3.03 Å, corresponding to second nearest neighbor separations. This indicates that the primary bonding takes place between Be and N atoms, while connection among similar atoms are largely responsible for maintaining the stability of the lattice structure. In addition, the Be-N-Be bond angle ($\theta_{\mathrm{Be-N-Be}}$), which is about 120°, reinforces a flat shape that aligns with the hexagonal symmetry of the monolayer.

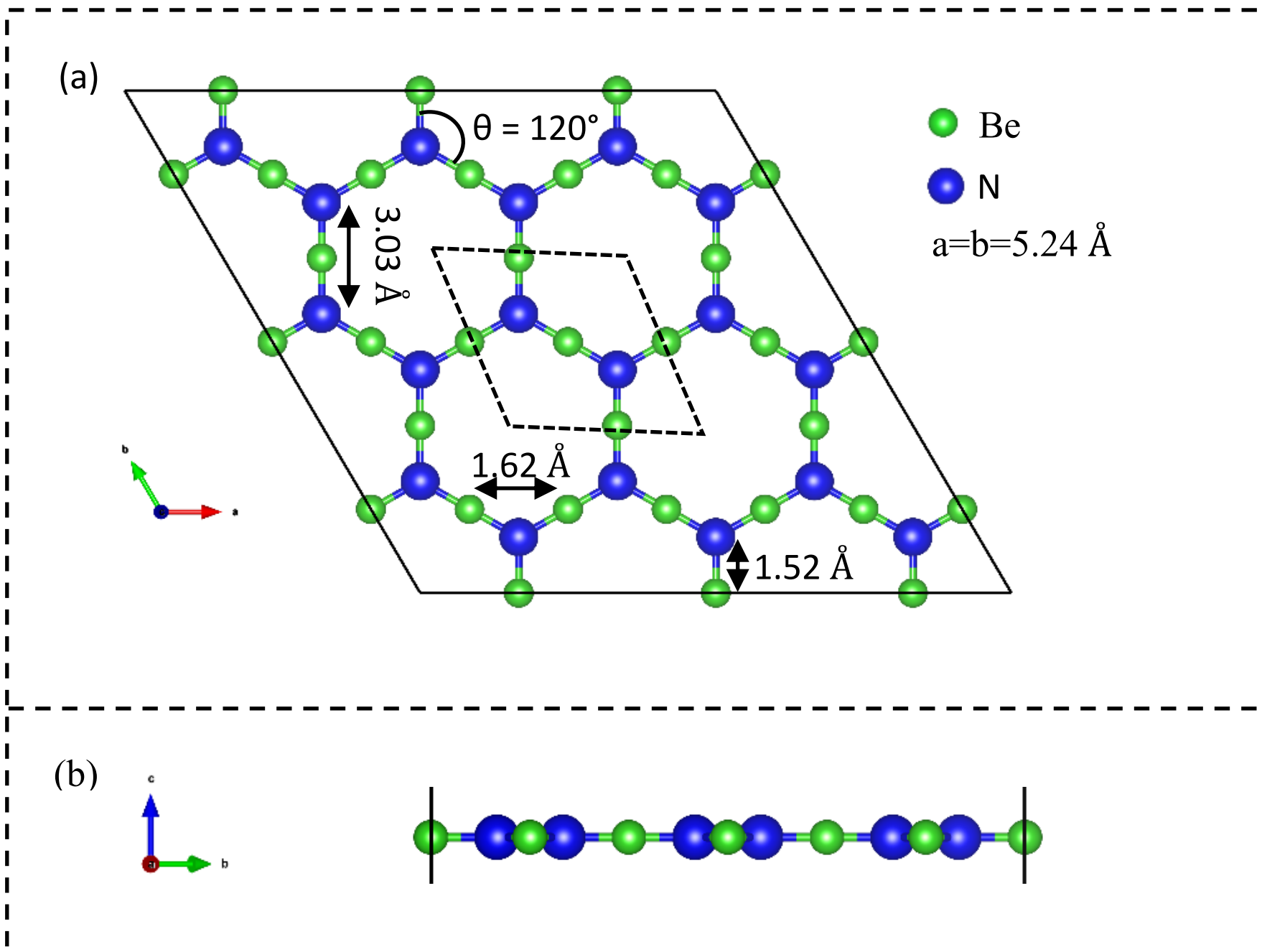

*Figure 1 (a) top, and (b) side view of the energy minimized crystal structure of $Be_3N_2$ monolayer with a primitive unit cell denoted by dashed line.*

Once the single layer structure has been effectively created, it is essential to evaluate its stability. Among the various influencing factors, cohesive and formation energies are often considering a key indicator of material's stability. Cohesive energy represents the energy required to break a material down into its individual atoms, serving as an effective indicator of the strength of atomic bonds. Mathematically, the cohesive energy per atom of $h$-$Be_3N_2$ is given by the following equation [46]

$$E_{cohesive} = \frac{3E_{Be}^{iso} + 2E_{N}^{iso} - E_{Be_3N_2}^{total}}{5} \tag{1}$$

Where, $E_{Be_3N_2}^{total}$, $E_{Be}^{iso}$, and $E_{N}^{iso}$ is the total energy of the $h$-$Be_3N_2$'s unit cell, and energy of individual (Be and N) atom in their isolated state, respectively. The calculated cohesive energy of $h$-$Be_3N_2$ is 4.35 eV per atom, indicating strong interatomic interactions and excellent structural stability.

In addition to cohesive energy, the formation energy insight into the thermodynamic viability of material production. It measures the energy change that occurs when a material forms its constituent atoms in their most stable bulk forms. The formation energy per atom is expressed as [47]

$$E_{formation} = \frac{E_{Be_3N_2}^{total} - 3E_{Be}^{bulk} - 2E_{N}^{bulk}}{5} \tag{2}$$

Here, $E_{Be}^{bulk}$ and $E_{N}^{bulk}$ represents the energies of atoms Be and N in their respective stable bulk phases. The calculated formation energy of $h$-$Be_3N_2$ is -0.18 eV per atom. The negative formation energy confirms that the monolayer is thermodynamically stable with respect to its constituent elements and suggests that its experimental synthesis is energetically favorable. Therefore, both the high cohesive energy and negative formation energy demonstrates the intrinsic stability and feasibility of $h$-$Be_3N_2$ as a two-dimensional material.

Additionally, the investigation of the phonon dispersion of $h$-$Be_3N_2$ was carried out by employing density functional perturbation theory (DFPT) alongside the highly symmetrical path M–K–Γ–M within the Brillouin zone, as illustrated in figure 2(a). The lack of imaginary phonon frequencies throughout the entire Brillouin zone affirms the structural dynamical stability. The unit cell contains five atoms, resulting in a complete set of fifteen vibrational branches, which consists of three acoustic and twelve optical branches. As shown in figure 2(a), the acoustic phonon branch spans a frequency range from 0 to 178 $cm^{-1}$, and are associated with the flexural out-of-plane (ZA), in plane transverse (TA), and in plane longitudinal (LA) vibrational mode. Similarly, the optical phonon branch, are divided into out-of-plane (ZO), in-plane transverse (TO), and in-plane longitudinal (LO) vibrational mode, cover a frequency range from 178 to 1400 $cm^{-1}$. Notably, phonon states are absent in the frequency interval of 520-650 $cm^{-1}$ and 900-1200 $cm^{-1}$, indicating the presence of additional phonon gaps. Also, we can notice that frequency of longitudinal modes is larger than the transverse modes. Near the gamma point, both TA and LA modes show linear dispersion, which is a characteristic commonly found in two-dimensional material. In contrast, the ZA mode deviates from ideal quadratic behaviour close to the Γ point, possibly due to a loss of reflection symmetry. Moreover, at Γ point, no splitting between LO and TO mode can be observed, which further supports the inherent 2D nature of the system.

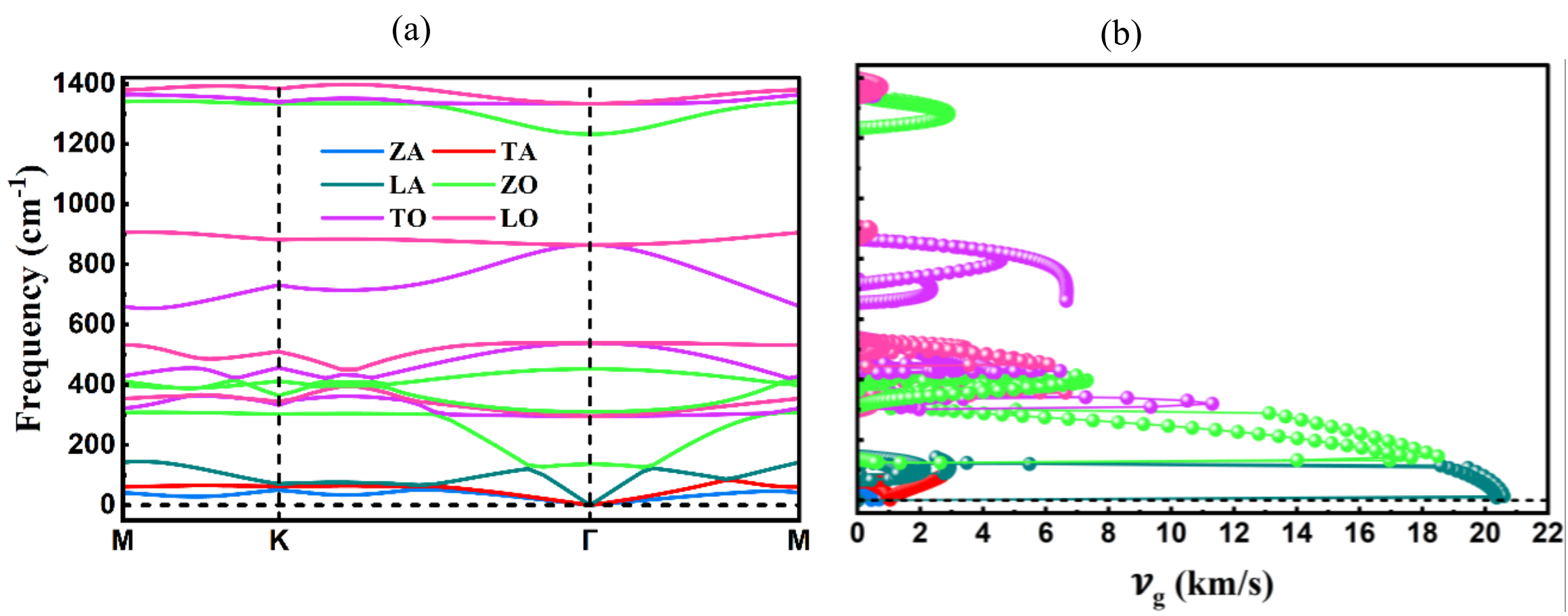

*Figure 2 (a) phonon dispersion curve, and (b) associated phonon group velocity ($v_g$) for the $Be_3N_2$ primitive unit cell, along the high symmetric direction (M–K–Γ–M) of Brillouin zone.*

The phonon group velocities, illustrated in figure 2(b), provided additional insight into the characteristics of thermal transport. Phonon group velocity was assessed by taking the derivative of the phonon angular frequency with respect to the wave vector, defined as ($v_g = d\omega/dk$), where $\omega$ is the phonon angular frequency and $k$ is the wave vector [48]. The group velocity plays an important role in the conduction of heat. As the frequency increases, the optical phonon branches become flatter, leading to lower group velocities and consequently a lesser contribution to heat transport. Among the acoustic modes, the group velocity of LA is higher than TA and ZA. In conclusion, the LA and ZO modes are anticipated to significantly influence thermal transport because of their comparatively high group velocities, estimated at approximately 18.38 km/s and 20.78 km/s, respectively.

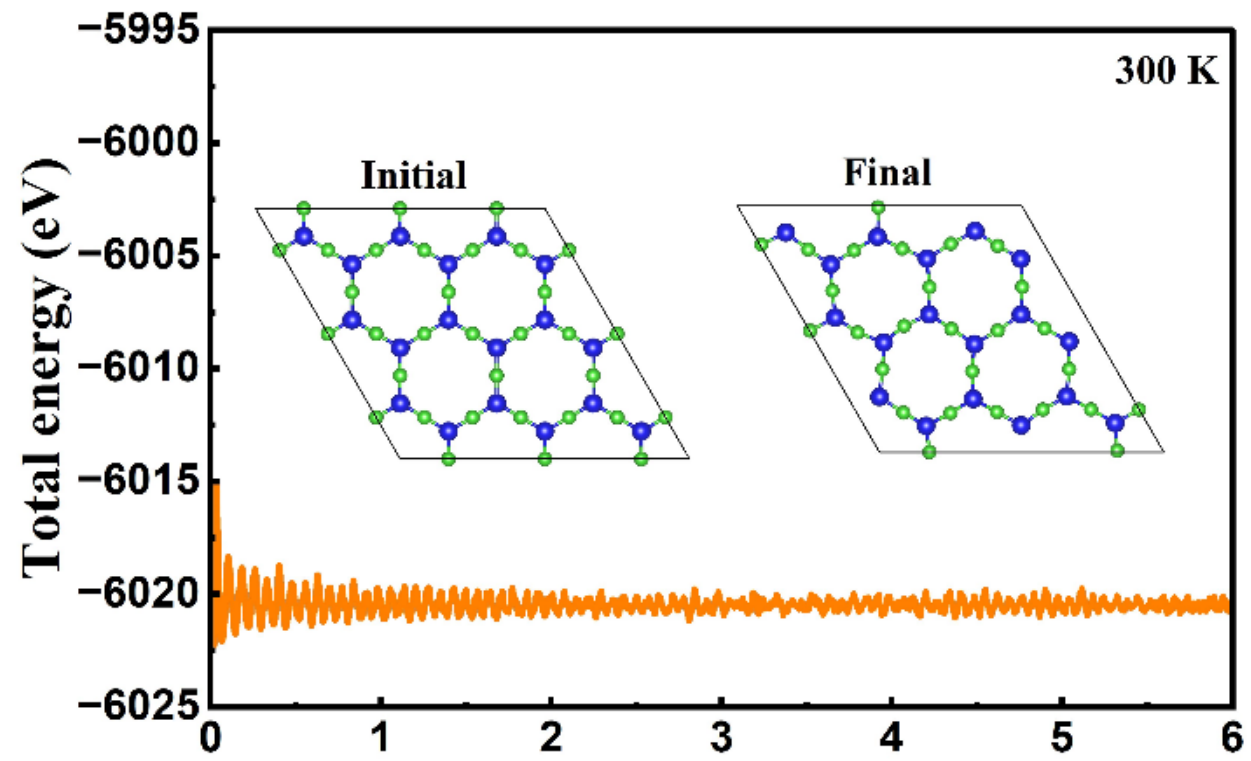

*Figure 3 total energy fluctuation of h-$Be_3N_2$ monolayer throughout an ab-initio molecular dynamics (AIMD) simulation carried out at 300 K for 6 ps.*

Moreover, to verify the thermal stability of $h$-$Be_3N_2$, ab-initio molecular dynamic (AIMD) simulation was performed within the NVT ensemble utilizing a Nosé-Hoover thermostat at a temperature of 300 K, based on the Car-Parrinello molecular dynamics (CPMD) method [49]. The system was simulated for a cumulative duration of 6 ps. As shown in figure 3, total energy fluctuation remains fairly constant around a fixed average value throughout the entire simulation period, suggesting that the system is well thermally equilibrated. We observed a small alteration in some of the bond lengths ($d_{\mathrm{Be-N}}$) and bond angle ($\theta_{\mathrm{Be\text{-}N\text{-}Be}}$) in the final atomic arrangement when compared to the initial structure. Nonetheless, this change is very small and can be attributed to natural atomic vibrations at 300 K. Crucially, there is no evidence of bond breaking or any distortion of the structure throughout the entire simulation process, indicating that $h$-$Be_3N_2$ effectively maintained under thermal circumstances.

## 3.2 Electronic properties

### 3.2.1 Band structure and projected density of states (PDOS)

After confirming both structural and dynamic stability of $h$-$Be_3N_2$, the next stage involved assessing its electronic characteristics to determine its suitability for photocatalytic applications. The analysis utilized the PBE exchange correlation functional to examine the electronic band structure and projected density of states (PDOS), offering an in-depth understanding of $h$-$Be_3N_2$'s electronic features and participation of its orbitals. The evolution of the band structure was conducted along the M-K-Γ-M high-symmetry trajectory, as depicted in figure 4(a). Findings indicate that $h$-$Be_3N_2$ acts as a direct band-gap semiconductor, with both the minimal of conduction band (CBM) and maximal of valence band (VBM) are located at the middle (Γ-point) of the Brillouin zone. The determined band-gap energy of 2.74 eV shows its excellent capacity to absorb visible light. The PDOS depicted in figure 4(b) further elucidates the electronic states. The conduction band is situated above the fermi energy ($E_f$), primarily consists of Be-*s*, Be-*p*, N-*s*, and N-*p* orbitals. Conversely, N-*p*, and N-*s* orbitals predominately impact the valance band close to the $E_f$. This advantageous arrangements of electronic states of the bands facilitates to reducing charge recombination and enhance the photocatalytic performance of $h$-$Be_3N_2$ for comprehensive water splitting process by facilitating effective separation and transport of electron-hole pairs generated by light.

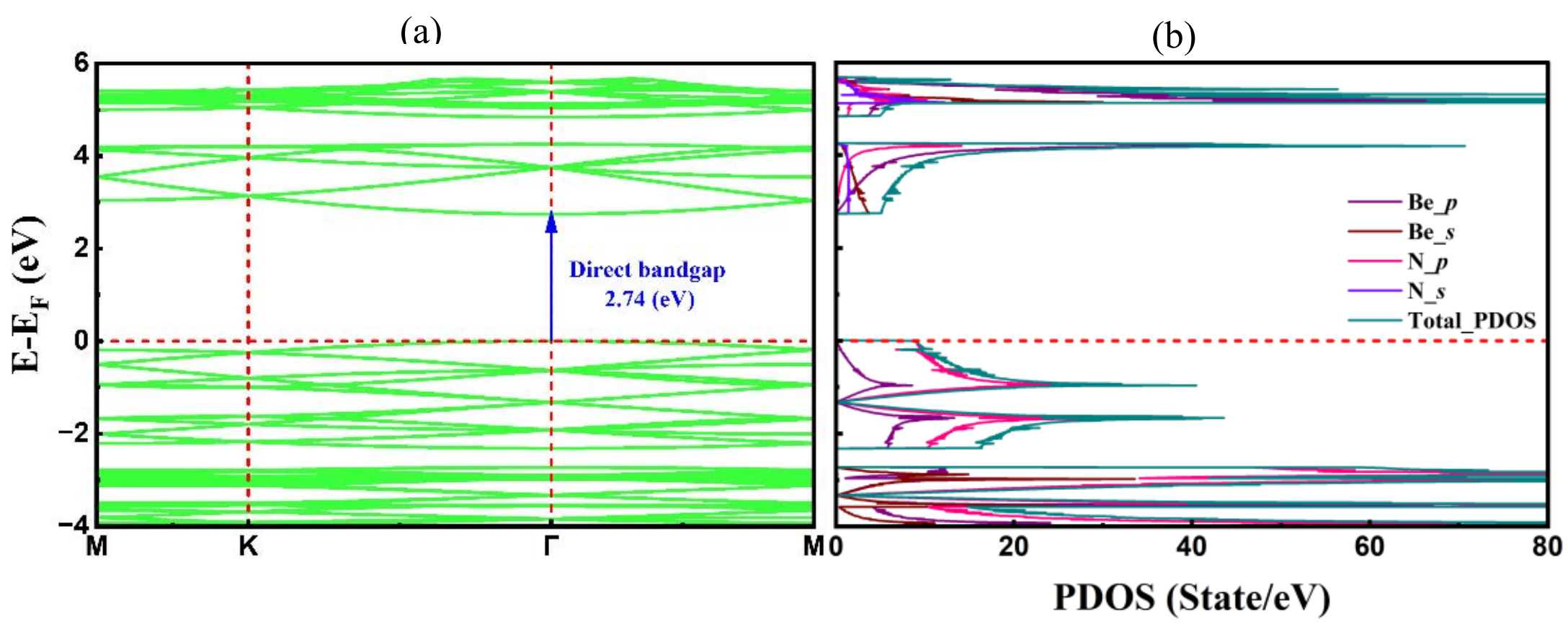

*Figure 4 (a) the electronic band structure of h-$Be_3N_2$ monolayer determined along the high symmetric path M-K-Γ-M within the Brillouin zone, along with (b) the relevant partial density of states (PDOS).*

### 3.2.2 Work function and Band edge alignment

To better understand the electron transfer kinetics, the work function of the $h$-$Be_3N_2$ was calculated. The work function is characterized as the lowest energy necessary to dislodge an electron from the surface of the material into the vacuum (or to the adsorbed intermediates), which can be represented by the equation

$$\Phi = E_{vacuum} - E_{fermi} \quad (3)$$

Where, $E_{vacuum}$ and $E_{fermi}$ indicates the energy of free electrons in free space and fermi energy, respectively [50,51]. These parameters were obtained from the electrostatic potential profile calculated along the z-axis, as shown in figure 5(a). It is noticed that the vacuum level energy of the $h$-$Be_3N_2$ is almost constant in vacuum due to symmetrical planer nature of $h$-$Be_3N_2$. The computed work function for $h$-$Be_3N_2$ is around 4.35 eV, which is lower compared to many earlier reported 2D photocatalyst, as shown in table 1. Although, this comparatively lower work function of $h$-$Be_3N_2$ suggests advantageous electron extraction and transfer characteristics, the overall effectiveness of photocatalysis cannot be evaluated solely on this work function.

*Table 1 the comparison of derived work function (Φ) of h-$Be_3N_2$ monolayer with those of recently reported 2D materials*

| **Materials** | **Work function** | **Reference** |
|---|---|---|
| $h$-$Be_3N_2$ | 4.35 | This work |
| $ZrSe_2$ | 5.47 | [52] |
| Pt | 5.56 | [53] |
| $Ge_2Se_2P_4$ | 5.85 | [54] |
| $SnSe_2$ | 5.91 | [55] |

The appropriateness for HER and OER is defined by the sites of the conduction band minimum (CBM) and the valence band maximum (VBM) relative to the redox potential of water splitting. The energies for CBM and VBM with respect to absolute vacuum scale (AVS) are calculated using the following equation:[56–58]

$$E_{CBM}^{AVS} = E_{CBM}^{DFT} - E_{vacuum} \tag{4}$$

$$E_{VBM}^{AVS} = E_{VBM}^{DFT} - E_{vacuum} \tag{5}$$

Where $E_{CBM}^{DFT}$, and $E_{VBM}^{DFT}$ are represents the actual energy of minimal conduction band and maximal valence band, respectively. Since, the water redox potential is depended on pH, the reduction potential $E_{H^+/H_2}^{AVS}$ and oxidation potential $E_{O_2/H_2O}^{AVS}$ as a function of pH with reference to AVS are calculated as [59,60]

$$E_{H^+/H_2}^{AVS} = -4.44 + pH \times 0.059\ eV \tag{6}$$

$$E_{O_2/H_2O}^{AVS} = -5.67 + pH \times 0.059\ eV \tag{7}$$

The parameters $E_{CBM}^{AVS}$, $E_{VBM}^{AVS}$, $E_{H^+/H_2}^{AVS}$, and $E_{O_2/H_2O}^{AVS}$ are then converted to the normal hydrogen scale (NHE) by utilizing the relation as given below,

$$E^{NHE} = -4.44\ eV - E^{AVS} \tag{8}$$

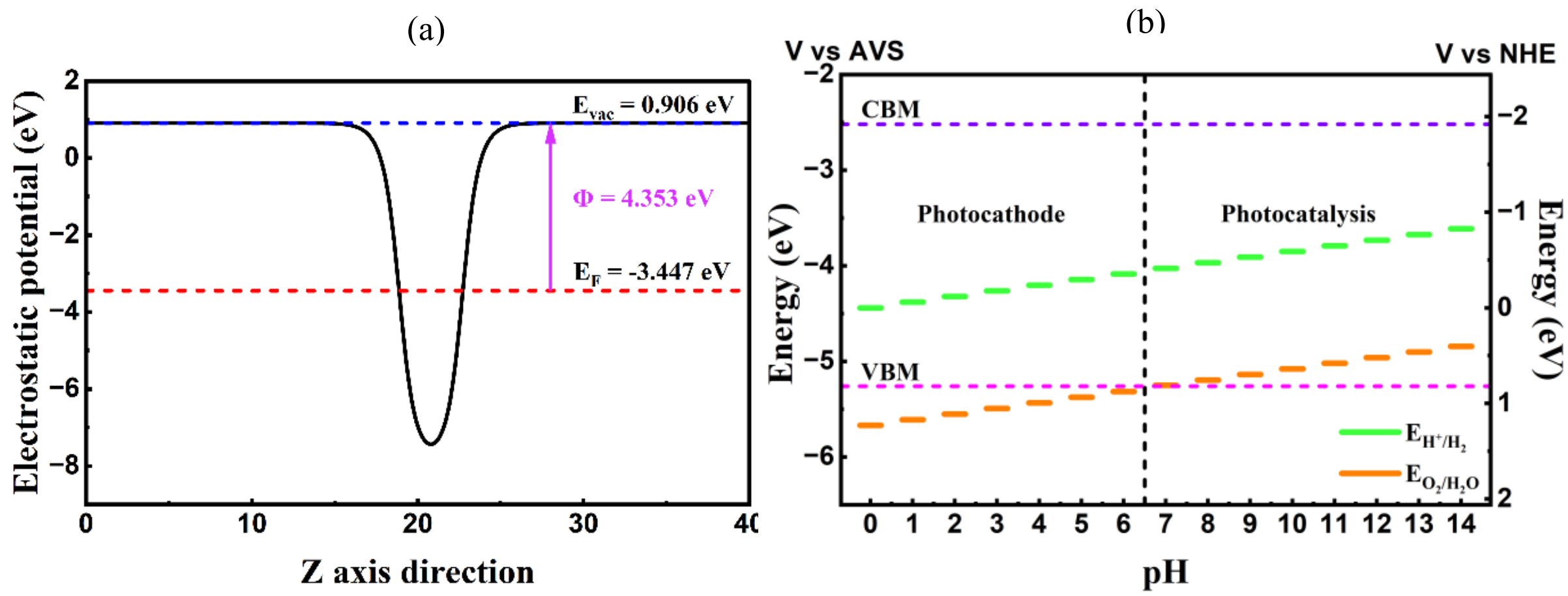

*Figure 5 (a) the average electrostatic potential of the h-$Be_3N_2$ monolayer perpendicular the z-axis (a), and (b) band-edge alignment in relation to AVS and NHE, accompanied by the reduction $E_{H^+/H_2}^{AVS}$ and oxidation potential $E_{O_2/H_2O}^{AVS}$ of water splitting, underscoring the potential of h-$Be_3N_2$ for HER and OER.*

The conduction and valence band-edge position of $h$-$Be_3N_2$ with respect to AVS and NHE scale, along with the water redox potential as a function pH are depicted in figure 5(b). The left vertical axis denotes the energy levels in relation to the AVS scale, while the right vertical axis is connected to the NHE scale. The location of CBM and VBM on AVS and NHE scale are found at approximately -2.51 eV (-1.94 eV) and -5.25 eV (0.94 eV) denoted by solid violet and magenta line. As shown in figure 5(a), the $E_{H^+/H_2}^{AVS}$ is denoted by dash green line, whereas the $E_{O_2/H_2O}^{AVS}$ is by orange, respectively. Furthermore, it can be noticed that the $E_{H^+/H_2}^{AVS}$ and $E_{O_2/H_2O}^{AVS}$ are increasing steadily with increasing the pH of water, which show the different nature of $h$-$Be_3N_2$ monolayer under different pH condition. Over the entire pH region ($0 \leq pH \leq 14$), the CBM stays far above the reduction potential $E_{H^+/H_2}^{AVS}$ of water, which confirms that the $h$-$Be_3N_2$ monolayer accurately satisfies the prescribed requirement of water reduction for the generation of hydrogen. Under the acidic region ($0 \leq pH \leq 6$), the VBM is higher compared to the $E_{O_2/H_2O}^{AVS}$, which indicates that $h$-$Be_3N_2$ monolayer doesn't meet the necessary criteria of water oxidation. Furthermore, at neutral pH (pH=7), the VBM is slightly less (approximately 0.1 eV) than the $E_{O_2/H_2O}^{AVS}$, suggesting that $h$-$Be_3N_2$ monolayer just fulfils the condition of oxidation of water. On the other hand, the VBM lies below the $E_{O_2/H_2O}^{AVS}$, confirming that $h$-$Be_3N_2$ monolayer adequately spans the water oxidation potential. This observation further indicates that, $h$-$Be_3N_2$ may be anticipated to act mainly as a photocathode under acidic media, since its band-edge position is incapable to promote oxygen during water splitting. Meanwhile, over the wide pH range (7≤pH≤14), $h$-$Be_3N_2$ could be used as an excellent photocatalyst, as its band-edge alignment effectively facilitates both HER and OER.

### 3.2.3 Charge density difference and Bader charge analysis

Before investigating the catalytic activity of HER and OER, we were analysed the difference of charge density to elucidate a deeper understanding of interfacial charge-transfer characteristics between the surface of $h$-$Be_3N_2$ and adsorbed reaction intermediates (H, OH, O and OOH). The evaluation of charge density change caused by the adsorption of H, OH, O, and OOH intermediates, was carried out using the following expression [61]

$$\Delta\rho = \rho_{Be_3N_2+intermediate} - \rho_{Be_3N_2} - \rho_{intermediate} \tag{9}$$

where $\rho_{Be_3N_2+intermediate}$ represents the electron charge density of $h$-$Be_3N_2$ with adsorption of intermediates, while $\rho_{Be_3N_2}$ and $\rho_{intermediate}$ denote the charge densities of the pristine $h$-$Be_3N_2$ surface and the isolated reaction intermediate, respectively, with all atomic positions fixed to those of the adsorbed configuration. Furthermore, the magnitude of charge transfer between the adsorbed intermediates and the $h$-$Be_3N_2$ surface were quantitatively determined using Bader charge analysis [62]. Figure 6(a-d) illustrates variation in the charge density difference of the completely optimized system, when the hydrogen (H), hydroxyl (OH), oxygen (O), and hydroperoxyl (OOH) adsorbed onto the $h$-$Be_3N_2$ monolayer, emphasizing favourable adsorption positions. The yellow area indicates where charge accumulates, whereas the cyan denotes regions of charge depletion.

For HER, we were attached single hydrogen atom at upper N site of $h$-$Be_3N_2$ monolayer. once the adsorption process was finished, the hydrogen accepted an electron from the surface. It is evident in the figure 6(a), where we can show that charge build-up is displayed around the hydrogen atom, resulting in an electron being lost from the surface. The amount of charge transfer is around 0.33 e, which is significant. In a meantime, during the OER process, all three intermediates (OH, O, and OOH) shows a tendency to be adsorbed at the hollow site situated at the centre of the hexagonal Be-N configuration. It can be observed from figures 6(b, c, and d) that the most significant charge transfer took place when O was adsorbed, in contrast to OH and OOH, respectively, indicating a sufficient electronic interaction between intermediates and *h*-$Be_3N_2$ monolayer. Furthermore, the most notable charge depletion occurred in the neighbouring hexagonal Be-N configuration during the adsorption. The charge transfer value for O is 1.62 e, while it measures 0.93 e for OH and 1.06 for OOH, notably. These findings revel that an adequate transfer of charge take place while adsorbing all intermediates (H, OH, O and OOH), which helps to facilitate both HER and OER.

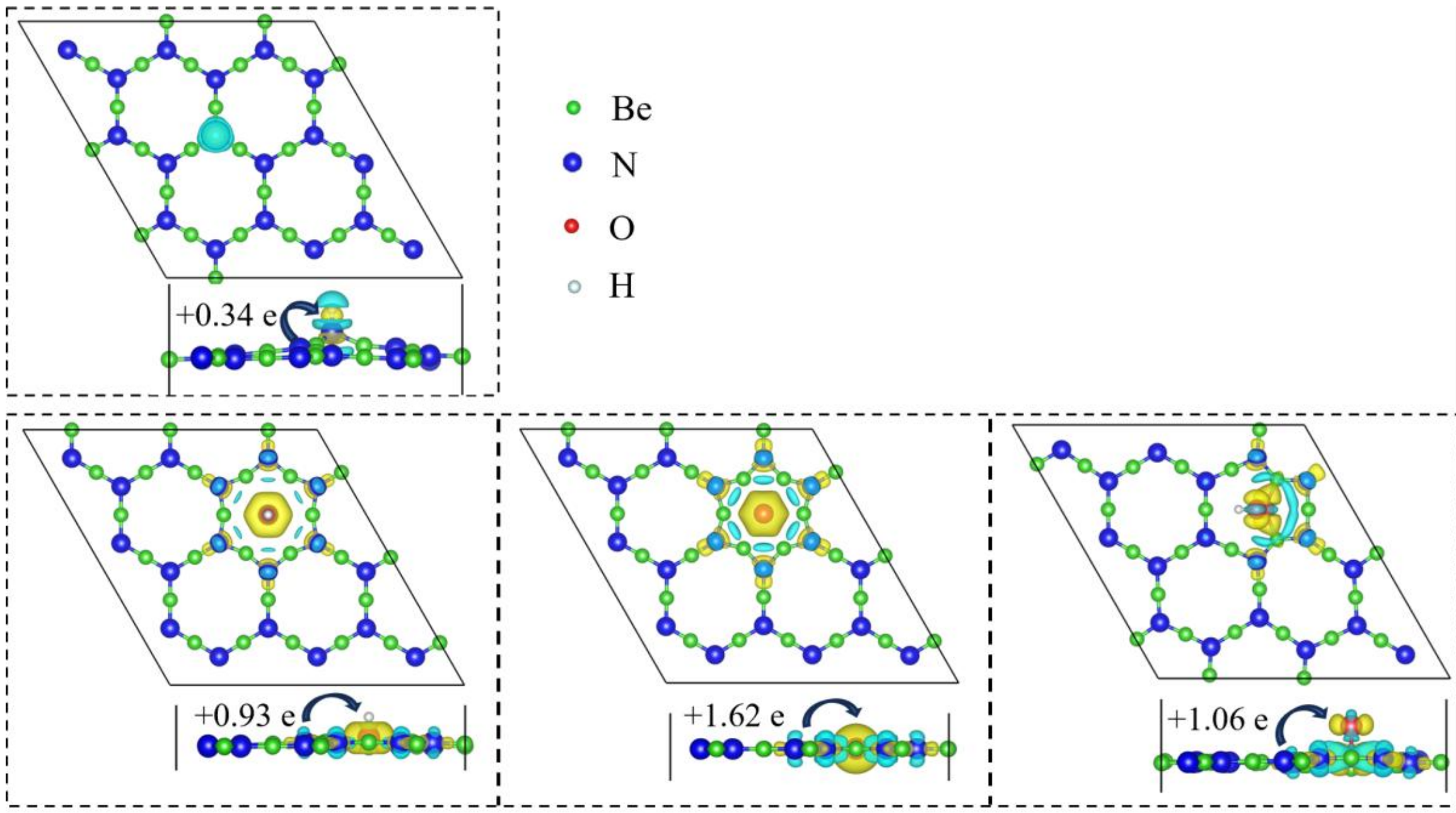

*Figure 6: (a-d) Charge density difference (CDD) iso-surfaces and corresponding Bader charge analysis for the adsorption of H*, OH*, O*, and OOH* intermediates on the h-Be3N2 monolayer. The CDD is visualized using iso-surface values of ±0.0018 e Å⁻³ for H*

## 3.3 Photocatalytic properties

### 3.3.1 Water adsorption and splitting

To investigate the photocatalytic capabilities of the *h*-$Be_3N_2$ monolayer on an atomic level, the initial research focused on how a water molecule interacts with the *h*-$Be_3N_2$ monolayer. This interaction is essential, as the water adsorption serves as a vital preliminary phase influencing the activation of the water molecule throughout the comprehensive process of water splitting. Before evaluating the catalytic activities associated with HER and OER, a suitable level of adsorption strength is required to support the next proton-coupled electron transfer reactions. The behaviour of adsorption was analysed utilizing the energy related with adsorption, which is expressed as [63]

$$E_{ads} = E_{H_2O^*} - E_* - E_{H_2O} \tag{10}$$

where $E_{H_2O^*}$, $E_*$, and $E_{H_2O}$ is the total energy of surface with adsorbed $H_2O$, pristine surface and $H_2O$ molecule, respectively. Initially, we placed a single water molecule above the Be and N atoms in the *h*-$Be_3N_2$ monolayer. Upon completing the optimization procedure, the $H_2O$ molecule moved toward the centre of the hexagon formed by Be and N, indicating that the optimal position for water adsorption is at the hollow site. The optimized configuration of *h*-$Be_3N_2$ with water adsorption is shown in figure 7(a). Furthermore, it is noticed that the bond angle of the H-O-H ($\theta_{H\text{-}O\text{-}H}$) was measured at 105°, and the O-H bond length was determined to be 1 Å, respectively. Additionally, the calculated adsorption energy of -0.30 eV suggests that water is spontaneously and stably held on the surface, without creating excessively strong bonds that could reduce the availability of *h*-$Be_3N_2$ monolayer's active sites. The determined energy of adsorption indicates that the water molecule engages sufficiently with the *h*-$Be_3N_2$ surface, which is absolutely necessary to achieve high catalytic efficiency throughout the whole water splitting.

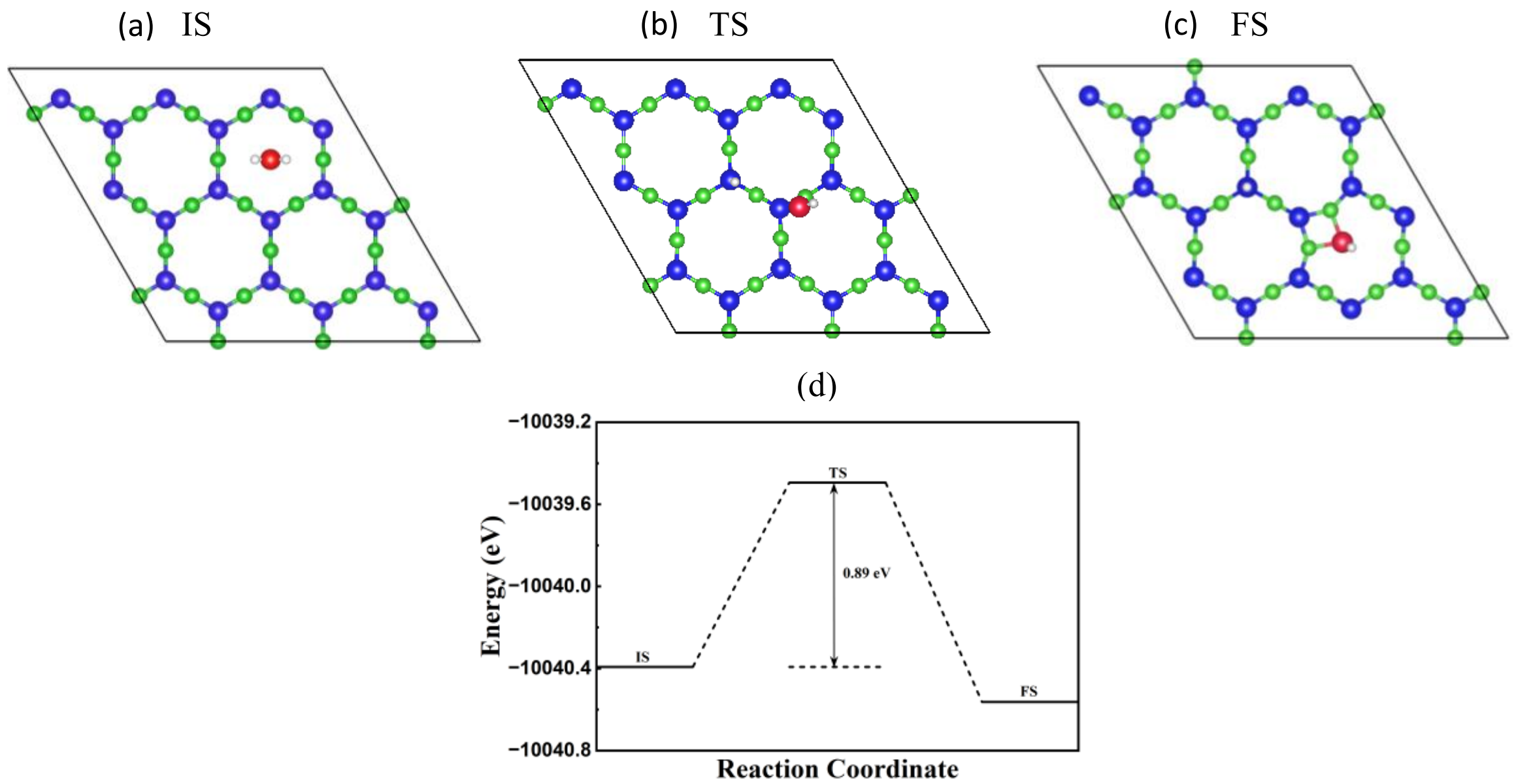

*Figure 7 (a-c) the optimized configuration for the initial state (IS), transition state (TS), and final state (FS) states for the most favourable H2O* → OH* + H* dissociation pathway reaction on h-$Be_3N_2$ monolayer, and (b)the associated minimal energy pathway and activation energy barrier for water dissociation derived from CI-NEB calculation, illustrating the kinetics of water activation on h-$Be_3N_2$ monolayer.*

After a successful adsorption of $H_2O$ molecule on $h$-$Be_3N_2$, we were further systematically explored pathways for the dissociation of water using the climbing-image nudged elastic band (CI-NEB) approach. In the process of splitting water mainly in basic media, the first step involves the breakdown of an adhered water molecule, which is essential for producing $H^+$ and $OH^-$ needed for the following HER and OER process. The reaction for water dissociation can be represented as[63]

$$H_2O^* \rightarrow H^* + OH^* \tag{11}$$

In which, * denotes the active site on the $h$-$Be_3N_2$ monolayer. The optimized geometries of the initial state (IS), transition state (TS), and final state (FS), together with the corresponding minimum-energy pathway obtained from the CI-NEB calculations [64], are illustrated in Figure 7(a-c). In this situation, it can be noted that the water molecule that is adsorbed at hollow site disintegrate into H and OH. Subsequently, the H binds to the top of the N atom, whereas the OH adsorbed beside hollow site once the procedure is finished. The activation energy was calculated based on the energy disparity between the TS and the IS, denoted as

$$E_{act} = E_{TS} - E_{IS} \tag{12}$$

while the reaction energy was determined using

$$\Delta E = E_{FS} - E_{IS} \tag{13}$$

where FS refers to the state after dissociation [63]. In equation 12 and 13, $E_{IS}$, $E_{TS}$, and $E_{FS}$ demonstrate the energy of IS, TS, and FS, respectively. The activation energy of water dissociation is computed around 0.89 eV, whereas the reaction energy is -0.17 eV. Although the dissociation requires overcoming a moderate kinetic barrier, the reaction is thermodynamically favourable with a negative reaction energy of −0.10 eV, indicating that the dissociated state is slightly more stable than the molecularly adsorbed state. The exothermic nature of the reaction suggests that once the TS is crossed, the generated H and OH intermediates can be stably accommodated on the surface, which is beneficial for subsequent HER and OER elementary steps. Overall, the combination of moderate water adsorption, a manageable dissociation barrier, and an exothermic reaction energy demonstrate that the $h$-$Be_3N_2$ possesses a stable and catalytically active surface for water activation, making it a promising candidate for not only acidic and neutral but also alkaline photocatalytic water splitting.

### 3.3.2 Gibb's free energy

We have shown that water molecules can be readily adsorbed and dissociated on the $h$-$Be_3N_2$ surface, our focus then turned to assessing the catalytic activity towards the HER and OER by developing the Gibb's free energy profile for the elementary reaction steps.

**Hydrogen evolution reaction:** For the photocatalytic reaction of hydrogen production, hydrogen can be generated through two main reaction routes, which is typically adheres to either the Volmer-Heyrovsky mechanism, which involves one adsorbed hydrogen intermediate along with a proton/electron transfer step, or the Volmer-Tafel mechanism, in which two adsorbed hydrogen atoms come together to produce $H_2$ molecule. The first step in the both routes is same, involves the adsorption of hydrogen onto the surface of h-$Be_3N_2$.

$$(H^+ + e^-) + * \rightarrow H^*(Volmer\ step) \tag{14}$$

Volmer step is then followed by either Heyrovsky step or Tafel step to form molecular hydrogen.

$$H^* + (H^+ + e^-) \rightarrow H_2 (Heyrovsky\ step) \quad (15)$$

or

$$H^* + H^* \rightarrow H_2 + * \ (Tafel\ step) \quad (16)$$

The Gibbs free energy chang associated for the hydrogen evolution reaction (HER) was evaluated utilizing the computational hydrogen electrode (CHE) framework introduced by Nørskov and his team mate, is given by the following equation [65,66]

$$\Delta G_{H^*} = \Delta E_{H^*} - T\Delta S_{H^*} + \Delta E_{ZPE} + \Delta G_U + \Delta G_{pH} \quad (17)$$

Where $\Delta E_{H^*}$ is the total energy of catalyst after hydrogen adsorption is determined as

$$\Delta E_{H^*} = E_{nH^*} - E_{(n-1)H^*} - \frac{1}{2} E_{H_2} \quad (18)$$

Where $E_{nH^*}$ and $E_{(n-1)H^*}$ demonstrate the total energy of *h*-$Be_3N_2$ after n hydrogen adsorption and (n-1) hydrogen coverage system, and $E_{H_2}$ is the energy of $H_2$ molecule at gas phase, respectively. Meanwhile, $\Delta S_{H^*}$ and $\Delta E_{ZPE}$ are the variation of entropy and zero-point energy. $\Delta S_{H^*}$ can be expressed as

$$T\Delta S_{H^*} = TS_{H^*} - \frac{1}{2} TS_0 \quad (19)$$

Where $TS_{H^*}$ is the vibrational entropy of *h*-$Be_3N_2$ with hydrogen adsorption, has been determined to be 0.026 eV, which was initially computed for a platinum surface at a temperature of 300 K [67]. As shown in equation (19), $TS_{H^*}$ is regarded as roughly equivalent for hydrogen adsorption across various types of material surfaces, and therefore, this same value is applied for the *h*-$Be_3N_2$ surface in this research. And $TS_0$ is approximately 0.41 eV according to NIST data [68]. Subsequently, $\Delta E_{ZPE}$ is the change in zero-point energy, formulated as

$$\Delta E_{ZPE} = E_{ZPE}^{nH^*} - E_{ZPE}^{(n-1)H^*} - \frac{1}{2} E_{ZPE}^{H_2} \quad (20)$$

Where $E_{ZPE}^{nH^*}$ and $E_{ZPE}^{H_2}$ is the zero point energy of n hydrogen atoms on the *h*-$Be_3N_2$ surface and single hydrogen molecule in the gas phase. The zero point energy correction for adsorbed intermediates can be computed from the vibrational frequency calculated using density functional perturbation theory (DFTP) [69].

$$E_{ZPE} = \frac{1}{2} h\nu_i \tag{21}$$

Where $\nu_i$ is the vibrational frequency. The vibrational frequency of adsorbed hydrogen on the $h$-$Be_3N_2$ are found to be 3035.4318 $cm^{-1}$, 834.9253 $cm^{-1}$ and 536.4 $cm^{-1}$, whereas 4463.4 $cm^{-1}$ for $H_2$ molecule, respectively. Now, substituting all this value in equation 1, we get the following equation

$$\Delta G_{H^*} = \Delta E_{H^*} + 0.31\ eV + \Delta G_U + \Delta G_{pH} \tag{22}$$

Furthermore, $\Delta G_U = -eU_e$, where $U_e$ is the electrode potential of the photogenerated electrons with respect to the AVS and e signifies the elementary charge. And $\Delta G_{\mathrm{pH}} = 0.059 \times \mathrm{pH}$. The potential of photogenerated electrons ($U_e$) is usually denoted as the difference between the CMB and reduction potential, given as below [66]

$$U_e = CBM - (-4.44) - 0.059 \times pH \tag{23}$$

In line with Sabater's principle, a photocatalyst optimized for the hydrogen evolution reaction (HER) needs to have a Gibbs free energy for hydrogen adsorption $\Delta G_{H^*}$ that is as close to zero as it can be. This guideline signifies a proper equilibrium between the adsorption and desorption of hydrogen on the surface of the catalyst, facilitating swift reaction kinetics and enhancing catalytic effectiveness. If the $\Delta G_{H^*}$value is positive, it denotes weak adsorption of hydrogen, rendering the initial step of proton adsorption thermodynamically unfavourable and thus constraining HER performance. On the other side, a strongly negative $\Delta G_{H^*}$ value is associated with excessively strong hydrogen attachment, which makes it difficult for $H_2$ molecules to desorb and hampers catalyst turnover. Hence, achieving a $\Delta G_{H^*}$ value that is close to zero (between -0.2 eV to +0.2 eV) is seen as a key characteristic of an ideal HER catalyst, as it guarantees a balanced process of adsorption and desorption and allows for effective hydrogen production [70].

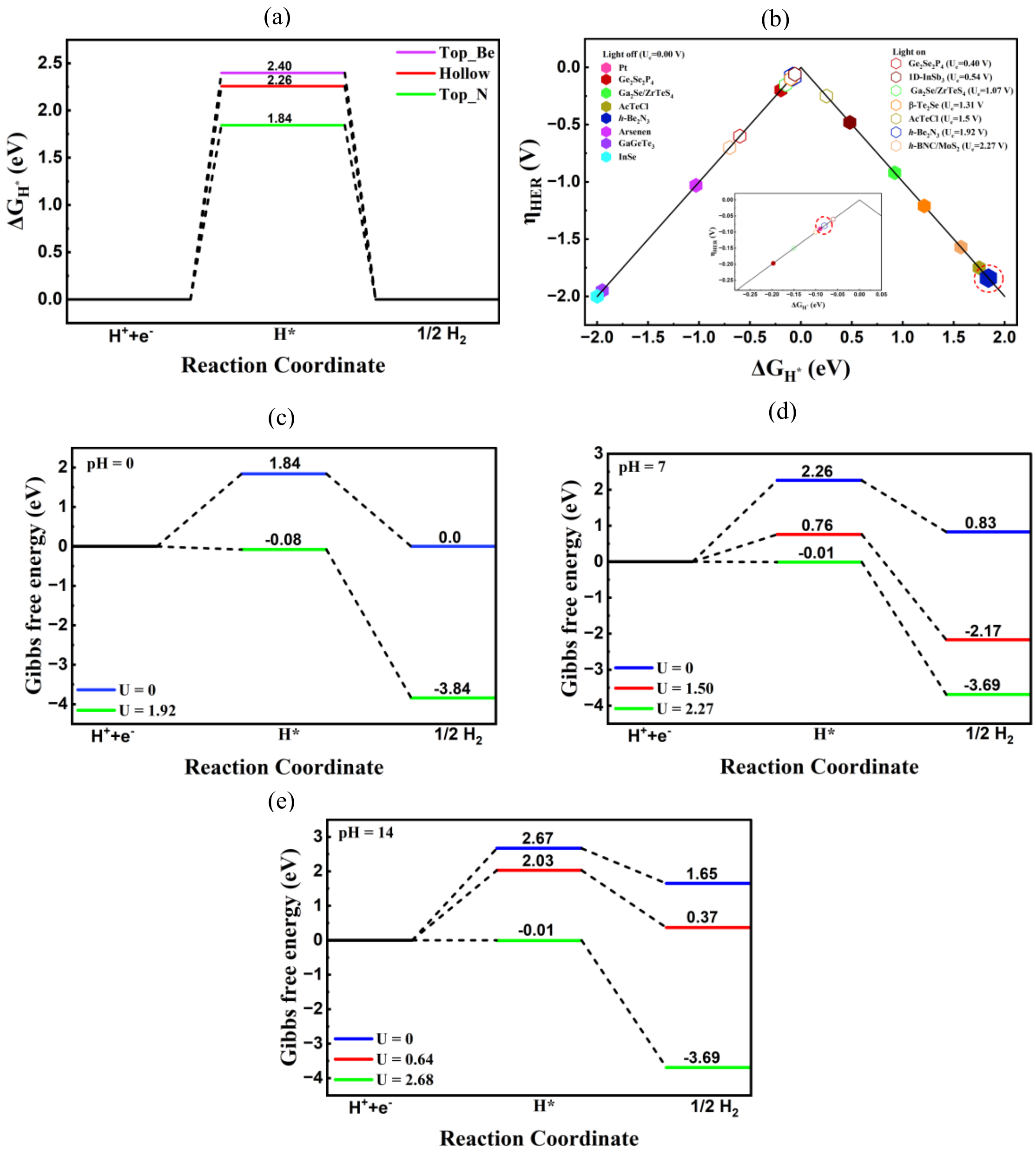

*Figure 8 (a) reaction coordinates for hydrogen adsorption at different sites on h-$Be_3N_2$ monolayer, (b) a comparison of the theoretical HER overpotential for the h-$Be_3N_2$ monolayer against recently reported 2D photocatalysts, (c) Gibbs free energy diagram for the HER on the h-$Be_3N_2$ monolayer under dark and illuminated scenario at (c) pH=0, (d) pH=7, and (e) pH=14.*

We determined the Gibbs free energy ($\Delta G_{H^*}$) associated with hydrogen adsorption at various sites on the $h$-$Be_3N_2$ monolayer. The adsorption locations on the $h$-$Be_3N_2$ monolayer include the top of the nitrogen atom, the top of the beryllium atom, and the centre of the hexagonal ring, which is marked by a cross in figure 1(a). Under standard conditions (U=0 V, and pH=0), the calculated free energy for adsorption at these diverse sites is illustrated in figure 8(a). The value of $\Delta G_{H^*}$ at the top nitrogen site is 1.84 eV, which is lower by 0.55 eV and 0.41 eV compared to the values at the top Be and hollow sites, respectively. Consequently, the top nitrogen site is recognized as the most favourable location for hydrogen adsorption on the $h$-$Be_3N_2$ monolayer and is anticipated to act as the main active site for the HER. Additionally, under dark condition, the $\Delta G_{H^*}$ is 2.26 eV and 2.84 eV at pH=7 and 14 as shown in figure 8(c) and 8(d). This notably high positive value (at pH=0,7 and 14) signifies that the creation of the adsorbed hydrogen intermediate (H*) via the step of proton-coupled electron transfer is endothermic. Therefore, the process of hydrogen adsorption on the $h$-$Be_3N_2$ surface faces significant barriers, leading to limited HER effectiveness without light.

The HER performance of the $h$-$Be_3N_2$ monolayer was evaluated and compared with several previously reported photocatalytic materials by calculating the overpotential at a pH of 0. Figure 8(b) illustrates the volcano plot associated with $\Delta G_{H^*}$ for the h-$Be_3N_2$ and various other recently reported materials. As shown in figure 8(b), it is noted that the 1D-$InSb_3$, GaSe/$ZrTeS_4$, and B-$Te_2Se$ show limited HER effectiveness under dark environment as their $\Delta G_{H^*}$ are 0.48 eV, 0.92 eV, and 1.21 eV. On the other hand, when exposed to light, all three photocatalysts exhibits outstanding HER activity, with $\Delta G_{H^*}$ values dropping to -0.06 eV, −0.15 eV and −0.10 eV, respectively, suggesting that photoexcitation greatly improves their ability to evolve hydrogen [66,71,72]. Therefore, we studied the HER performance of $h$-$Be_3N_2$ under light on condition at different pH value, presented in figure 8(c) to 8(e). Under light irradiance at pH=0, the electrons generated by photons create a large amount (1.92 V) of potential internally, which substantially reduces the $\Delta G_{H^*}$ from 1.84 eV to -0.08 eV. This result is nearly aligned with the ideal thermoneutral state ($\Delta G_{H^*} = 0\ eV$), reflecting a phenomenal photocatalytic performance of the $h$-$Be_3N_2$ monolayer in illuminated condition without required any additional potential input in acidic media.

Moreover, when the light was illuminated, the electrons generated by light create internal potentials of 1.50 V (pH=7) and 0.64 V (pH=14). Under the influence of the photogenerated potentials the $\Delta G_{H^*}$ is decreases to 0.76 eV (pH=7) and 2.03 eV (pH=14). Although, the $\Delta G_{H^*}$ values are still above the optimal value ($\Delta G_{H^*}$=0 eV), indicating that spontaneous hydrogen evolution is not thermodynamically possible even in light irradiance. In the neutral medium (pH = 7), an external potential of 0.77 V is essential to promote the HER in an exothermic manner. This potential requirement is lower than that needed for $SiP_2$, PE-$AgBiP_2Se_6$, and FE-$AgBiP_2Se_6$, which are approximately 0.83 eV, 1.06 eV, and 1.62, highlighting the superior performance of the $h$-$Be_3N_2$ monolayer regarding HER [74,75]. Moreover, in alkaline environments, the energy barrier is substantially elevated, necessitating a considerably higher external potential to initiate hydrogen production. Consequently, despite the considerable enhancement brought about by light exposure, the $h$-$Be_3N_2$ shows outstanding hydrogen evolution reaction performance primarily in acidic environments, whereas its effectiveness diminishes in neutral and is thermodynamically unfavourable in basic environments.

**Oxygen evolution reaction:**

Following the validation of the HER catalytic activity of the $h$-$Be_3N_2$ monolayer, we systematically evaluated its performance towards the OER determine its viability for overall photocatalytic water splitting. Typically, the OER occurs through a sequence of four proton-coupled electron transfer steps, leading to the creation of essential adsorbed intermediates on the surface of the catalyst: OH*, O*, and OOH*, where * signifies the active catalytic location. The fundamental reaction pathways can be represented as follows [72]:

$$* + H_2O \rightarrow OH^* + H^+ + e^- \quad (24)$$

$$OH^* \rightarrow O^* + H^+ + e^- \quad (25)$$

$$O^* + H_2O \rightarrow OOH^* + H^+ + e^- \quad (26)$$

$$OOH^* \rightarrow * + O_2 + H^+ + e^- \quad (27)$$

The change in Gibbs free energy ($\Delta G_1$, $\Delta G_2$, $\Delta G_3$, and $\Delta G_4$) for these consecutive elementary steps were calculated as:

$$\Delta G_1 = G_{OH^*} + G_{(H^+ + e^-)} - G_* - G_{H_2O} + \Delta G_U - \Delta G_{pH} \tag{28}$$

$$\Delta G_2 = G_{O^*} + G_{(H^+ + e^-)} - G_{OH^*} + \Delta G_U - \Delta G_{pH} \tag{29}$$

$$\Delta G_3 = G_{OOH^*} + G_{(H^+ + e^-)} - G_{O^*} - G_{H_2O} + \Delta G_U - \Delta G_{pH} \tag{30}$$

$$\Delta G_4 = G_* + G_{(H^+ + e^-)} + G_{O_2} - G_{OOH^*} + \Delta G_U - \Delta G_{pH} \tag{31}$$

Here, $G_*$, $G_{OH^*}$, $G_{O^*}$, *and* $G_{OOH^*}$ refer to the Gibbs free energies associated with the clean surface and the respective adsorbed intermediates, as shown in table (2). The term $\Delta G_U = -eU_h$ considers the influence of the applied voltage, while $\Delta G_{pH} = 0.059 \times pH$ involves the correction for pH. The photogenerated hole potential ($U_h$) is defined as the energy difference between the reduction potential and the VBM, given as below [66]

$$U_h = (-4.44) - VBM - 0.059 \times pH$$

The equilibrium thermodynamic potential for the OER is established at 1.23 V, which corresponds to a total Gibbs free-energy change of 4.92 eV necessary for the oxidation of water involving four electrons. The catalytic capability of the $h$-$Be_3N_2$ monolayer for the OER was assessed by determining the theoretical overpotential $\eta_{OER}$, defined as [54]:

$$\eta_{OER} = max\left\{\frac{\Delta G_2, \Delta G_3, \Delta G_4, \Delta G_5}{e}\right\} - 1.23 \tag{32}$$

The step that shows the greatest change in Gibbs free energy indicates the minimum external potential needed to facilitate the reaction and is thus called the potential-determining step (PDS). A lower OER overpotential is indicative of better catalytic performance, suggesting more favourable energetic conditions for the reaction and improved efficiency in water oxidation. Therefore, the derived free-energy profile offers valuable insight into the OER process and allows for a quantitative evaluation of the photocatalytic performance in water splitting of the $h$-$Be_3N_2$ monolayer.

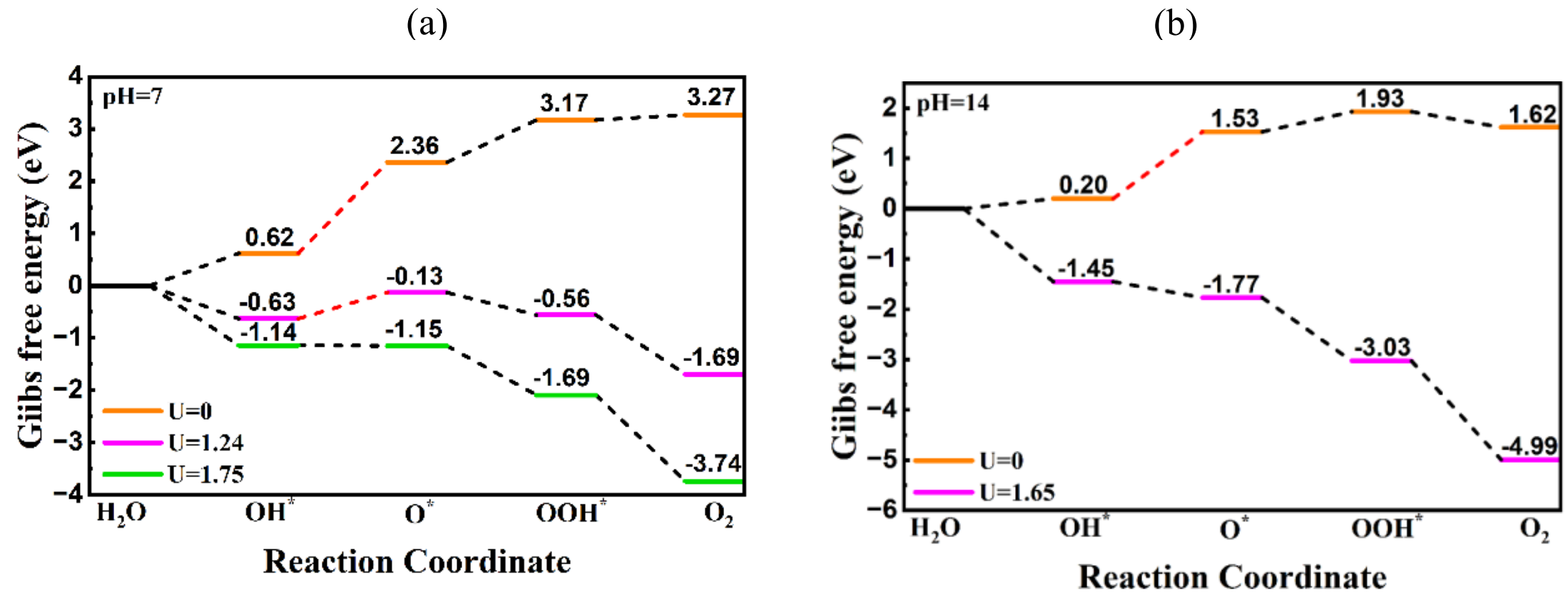

*Figure 9 Gibbs free energy profile for the OER on the h-$Be_3N_2$ monolayer under light on and light off environment at (a) pH=7 and (b) pH=14.*

As we can be seen in section of band-edge alignment, it was determined that the $h$-$Be_3N_2$ cannot effectively initiate the OER in acidic environments since its VBM does not yield adequate oxidation potential. Consequently, the Gibbs free energy change for the OER on $h$-$Be_3N_2$ monolayer was conducted exclusively at pH levels of 7 and 14, as illustrated in figure 10(a) and (b), respectively. As shown in figure 10 (a), without light at pH=7, the calculated changes in Gibbs free energy ($\Delta G_1$, $\Delta G_2$, $\Delta G_3$, and $\Delta G_4$) associated with four fundamental proton-coupled electron transfer step are calculated to be 0.62 eV, 1.74 eV, 0.81 eV, and 0.10 eV. Among these, the conversion of OH* to O* exhibits the highest free energy barrier ($\Delta G_2$), thus recognized as the PDS denoted by red line in figure 10 (a). The overall upward trend in free energy profile signifies that the OER mechanism is unspontaneous under dark environment. Although, the internal potential of 1.24 V produced by photoexcited holes, three of the four reaction steps become exothermic, the OH*→O* transformation is still remains endergonic. In order to surpass this thermodynamic limitation and enable the OER to occur spontaneously, the $h$-$Be_3N_2$ monolayer needs an external potential of at least 1.75 V. Taking into account the inherent photogenerated potential of 1.24 V, an extra potential of 0.51 V is necessary to overcome the remaining energy threshold and promote effective OER performance. The required extra potential is lower than the previously reported 2D material such as AsTeBr (1.98), GaAs (1.39 eV), g-$C_3N_4$ (1.45 eV), CuCl (1.53 eV), β-$Te_2S$ (1.25) etc. [72,73,76–78].

Additionally, as illustrated in Figure 9(b), the OER process is still thermodynamically unfavourable in dark environments at a pH of 14, as indicated by the positive changes in $\Delta G_1$, $\Delta G_2$, $\Delta G_3$, and $\Delta G_4$ throughout the reaction pathway. Nonetheless, the energy barrier for the PDS is lowered to 1.33 eV, which is an improvement compared to the value at pH of 7, signifying enhanced OER kinetics in alkaline conditions. When light is applied, the potential produced by the photogenerated holes greatly decreases the Gibbs free energies for all fundamental reaction steps. As a result, all fundamental reaction stages become thermodynamically advantageous, showing negative changes in $\Delta G_1$, $\Delta G_2$, $\Delta G_3$, and $\Delta G_4$ when exposed to light. This means that OER can occur on its own without needing any additional potential. These results indicate that the $h$-$Be_3N_2$ monolayer shows remarkable OER performance under alkaline circumstances when subjected to light exposure. The complete removal of uphill reaction barriers implies that photogenerated holes deliver an adequate potential to effectively facilitate the OER mechanism, confirming the promising capability of $h$-$Be_3N_2$ as a photocatalyst for solar-driven water oxidation in alkaline environments.

# 4 Effect of hydrogen coverage

Furthermore, we investigated the hydrogen evolution reaction abilities of $h$-$Be_3N_2$ by systematically increasing the coverage of hydrogen on its surface and evaluating the corresponding $\Delta G_{H^*}$ for each coverage increment in the presence of light at both pH levels of 0 and 7, as illustrated in figures 1 (a) & (b). In the 3×3×1 $h$-$Be_3N_2$ monolayer structure, there are 18 nitrogen atoms, resulting in 18 equivalent binding sites available for hydrogen atom adsorption on the surface. As a result, the H coverage is expressed as $\theta = 1/n$, where n can be 1, 2, …, or 18. An ideal range for $\Delta G_{H^*}$ is found between -0.20 and 0.20 eV, as indicated by the light green area, showing a favourable equilibrium between hydrogen adsorption and desorption at the surface of h-Be3N2. Figures 1(a) and 1(b) indicate that $\Delta G_{H^*}$ experiences considerable changes as the H coverage rises. Under specific conditions of pH=0 and U=1.92 V, the measured $\Delta G_{H^*}$ values are -0.08 eV, 0 eV, and -0.13 eV for coverage levels of 1/18, 7/18, and 13/18, respectively, suggesting that the HER is favourable at these specific coverage levels. Additionally, at coverage levels of 5/18 and 17/18, the HER is classified as moderate, with measured $\Delta G_{H^*}$ values of -0.22 eV and - 0.30 eV, respectively. Moreover, at other hydrogen coverage levels, $\Delta G_{H^*}$ values become excessively negative, falling between -0.5 and -1.38 eV, indicating difficulties in desorbing hydrogen.

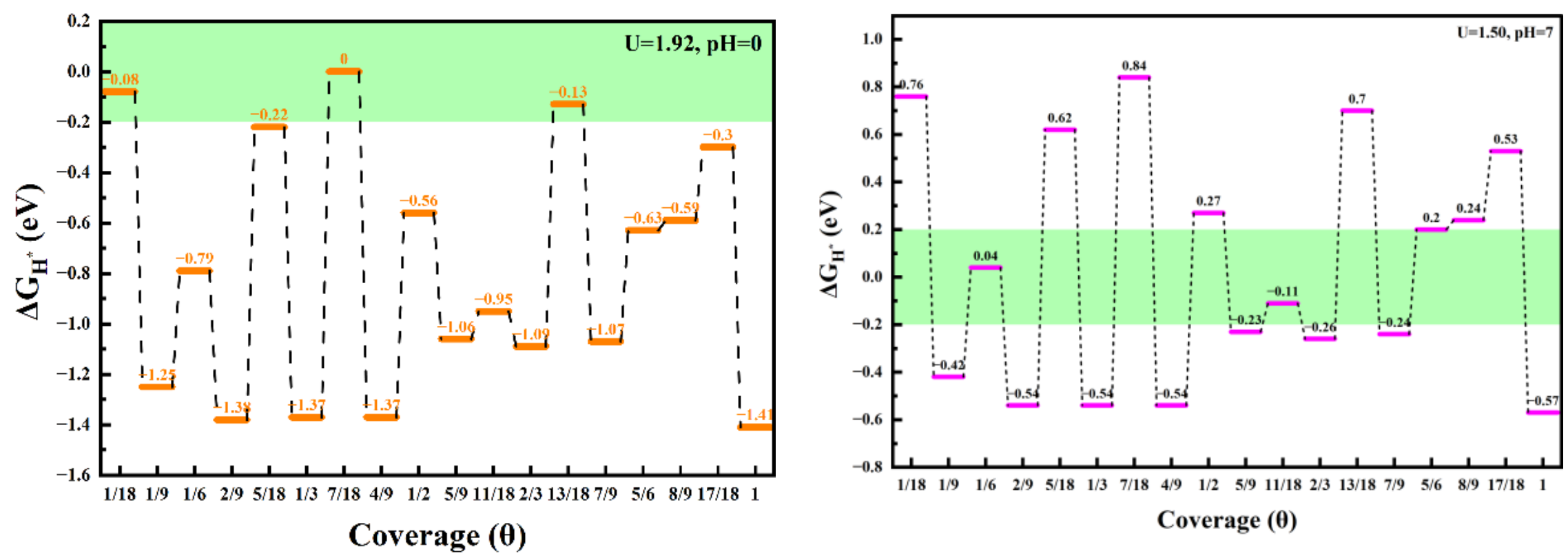

*Figure 10 hydrogen coverage in the presence of light at (a) pH=0, and (b) pH=7*

In contrast, when pH is 7 and U is 1.50 V, a different trend in coverage is observed. Here, $\Delta G_{H^*}$ dramatically fluctuates between the 0.84 eV to -0.57 eV. The HER is more favourable at coverage 1/6 and 11/18, at which $\Delta G_{H^*}$ values are 0.04 eV and -0.11 eV, notably. At coverage levels of 1/2, 5/9, 2/3, 7/9, and 8/9, the $\Delta G_{H^*}$ values are between 0.27 and -0.27 eV, correspondingly. At 1/2, 5/9, 2/9, 1/3, 4/9, and 1, the $\Delta G_{H^*}$ are roughly -0.5 eV, showing a moderate HER. Furthermore, the positive $\Delta G_{H^*}$ values (0.76 eV, 0.84, and 0.7 eV) are noticed at coverage levels of 1/18, 7/18, and 13/18 suggesting that binding of hydrogen is inadequate. The analysis focusing on coverage shows that the HER effectiveness of h-$Be_3N_2$ is significantly affected by the quantity of hydrogen that is adsorbed. The most advantageous thermodynamic condition for HER occurs at specific H coverage levels where $\Delta G_{H^*}$ nears zero, underlining the necessity of a balanced condition between the adsorption and desorption of hydrogen for optimal hydrogen evolution.

# 5 Conclusions

In this work, we have systematically investigated the structural stability, electronic properties, and photocatalytic performance of the 2D *h*-$Be_3N_2$ monolayer for water splitting application, utilizing a first-principles density functional theory. The calculated energetic, dynamic, and thermal properties confirm the excellent stability of $h$-$Be_3N_2$ monolayer, indicating its suitability for practical applications. Electronic structure calculations based on the PBE functional reveal that $h$-$Be_3N_2$ is a direct band-gap semiconductor with a band gap value of 1.74 eV at the Γ point, making it well suited for visible-light-driven photocatalysis. Furthermore, the alignment of band-edges illustrates that $h$-$Be_3N_2$ has suitable redox potential that facilitate overall photocatalytic water splitting in neutral and alkaline environment. Conversely, in acidic condition, the position of valence band is an adequate to enable the OER, thereby leading to photocathodic behavior.

To analyse its photocatalytic efficiency further, the Gibbs free energy alterations related to HER and OER were thoroughly assessed across various pH levels under both dark and illuminated circumstances. Findings indicate that when exposed to light, the $h$-$Be_3N_2$ monolayer shows nearly thermoneutral Gibbs free energy for HER in acidic settings (pH = 0), suggesting highly effective hydrogen production. At neutral pH, the HER process is still possible with the application of an external potential of 0.77 V. For the OER process, it is thermodynamically favourable only under strongly alkaline conditions, while an external potential of 0.43 V is needed to promote the reaction at neutral pH. These results clearly underscore the significant impact of solution pH and photogenerated charge carriers on the photocatalytic effectiveness of the $h$-$Be_3N_2$ monolayer. Furthermore, the study of H coverage indicates that the amount of H that adheres to the surface has a considerable impact on the efficiency of the HER in $h$-$Be_3N_2$. At a pH of 0, a coverage of 7/8 and at a pH of 7, a coverage of 11/18, h-Be3N2 exhibits remarkable HER since its $\Delta G_{H^*}$ value is 0 eV and -0.11 eV.

In general, the exceptional structural stability, advantageous electronic features, appropriate band-edge alignment, and favourable thermodynamic properties for both HER and OER underscore the considerable potential of the $h$-$Be_3N_2$ monolayer as an effective photocatalyst for solar-driven water splitting. This investigation provides insightful theoretical perspectives on Be-based two-dimensional materials and establishes a robust foundation for the intelligent design and advancement of future photocatalysts aimed at sustainable hydrogen production. Moreover, the favourable photocatalytic capabilities of the $h$-$Be_3N_2$ monolayer imply that it might be a viable option for upcoming solar-powered green hydrogen generation and may assist in realizing the goals of the National Green Hydrogen Mission by facilitating effective and sustainable hydrogen production.